\def \SAIT #1 #2 {{\em Mem.\ Soc.\ Astron.\ It.\/} {\bf #1}, #2}
\def \MESS #1 #2 {{\em The Messenger\/} {\bf #1}, #2}
\def \ASTRNACH #1 #2 {{\em Astron. Nach.\/} {\bf #1}, #2}
\def \AAP #1 #2 {{\em Astron. Astrophys.\/} {\bf #1}, #2}
\def \AAL #1 #2 {{\em Astron. Astrophys. Lett.\/} {\bf #1}, L#2}
\def \AAR #1 #2 {{\em Astron. Astrophys. Rev.\/} {\bf #1}, #2}
\def \AAS #1 #2 {{\em Astron. Astrophys. Suppl. Ser.\/} {\bf #1}, #2}
\def \AJ #1 #2 {{\em Astron. J.\/} {\bf #1}, #2}
\def \ANNREV #1 #2 {{\em Ann. Rev. Astron. Astrophys.\/} {\bf #1}, #2}
\def \APJ #1 #2 {{\em Astrophys. J.\/} {\bf #1}, #2}
\def \APJL #1 #2 {{\em Astrophys. J. Lett.\/} {\bf #1}, L#2}
\def \APJS #1 #2 {{\em Astrophys. J. Suppl.\/} {\bf #1}, #2}
\def \APSS #1 #2 {{\em Astrophys. Space Sci.\/} {\bf #1}, #2}
\def \ASR #1 #2 {{\em Adv. Space Res.\/} {\bf #1}, #2}
\def \BAIC #1 #2 {{\em Bull. Astron. Inst. Czechosl.\/} {\bf #1}, #2}
\def \JSQRT #1 #2 {{\em J. Quant. Spectrosc. Radiat. Transfer\/} {\bf #1}, #2}
\def \MN #1 #2 {{\em Mon. Not. R. Astr. Soc.\/} {\bf #1}, #2}
\def \MEM #1 #2 {{\em Mem. R. Astr. Soc.\/} {\bf #1}, #2}
\def \PLR #1 #2 {{\em Phys. Lett. Rev.\/} {\bf #1}, #2}
\def \PASJ #1 #2 {{\em Publ. Astron. Soc. Japan\/} {\bf #1}, #2}
\def \PASP #1 #2 {{\em Publ. Astr. Soc. Pacific\/} {\bf #1}, #2}
\def \NAT #1 #2 {{\em Nature\/} {\bf #1}, #2}

\def\g320{G320.4--01.2}
\def\msh{MSH~15--5{\em 2}}
\def\rcw{RCW~89}
\def\psr{B1509--58}
\def\HI{H\,{\sc i}}
\def\la{\ifmmode\stackrel{<}{_{\sim}}\else$\stackrel{<}{_{\sim}}$\fi} 

\documentstyle[epsfig]{memsait}
\begin{opening}
\title{RADIO OBSERVATIONS OF \g320\ AND PSR~\psr}
\author{B. M. GAENSLER$^{1,2}$\thanks{Current address:
Center for Space Research, Massachusetts Institute of Technology,
MA, USA. Email: bmg@space.mit.edu}, 
K. T. S. BRAZIER$^3$,
R. N. MANCHESTER$^2$, S. JOHNSTON$^4$, A.~J.~GREEN$^1$}
\institute{$^1$Astrophysics Department, School of Physics, University
of Sydney, NSW, Australia \\
$^2$Australia Telescope National Facility, CSIRO, NSW, Australia \\
$^3$Department of Physics, University of Durham, Durham, United Kingdom \\
$^4$Research Centre for Theoretical Astrophysics, University of Sydney,
NSW, Australia}
\date{} 
\end{opening}

\begin{document}

\oddpagefooter{}{}{} 
\evenpagefooter{}{}{} 
\ 
\bigskip

\begin{abstract} 

\g320\ is a complex radio and X-ray source, coinciding on
the sky with the young energetic pulsar \psr. A young pulsar
embedded in a SNR would seem to accord with expectations, but previous
observations suggest that all may not be what it seems. Controversy
persists over whether the pulsar and the SNR are associated, and as to what
causes the remnant's unusual appearance.  To answer these questions, we
have undertaken a set of high-resolution radio observations of the
system. We present the results of this study, which provide new
evidence that PSR~\psr\ is associated with and is interacting with
\g320.

\end{abstract}

\section{Introduction}

\g320\ (\msh) and the young pulsar PSR~\psr\ are one of the
best-studied but least-understood pulsar / supernova remnant (SNR)
pairings. We here summarise the results of a new radio study of this
system; this work is discussed in more detail by Gaensler et al
(1998).

\begin{figure}
\centerline{\epsfig{file=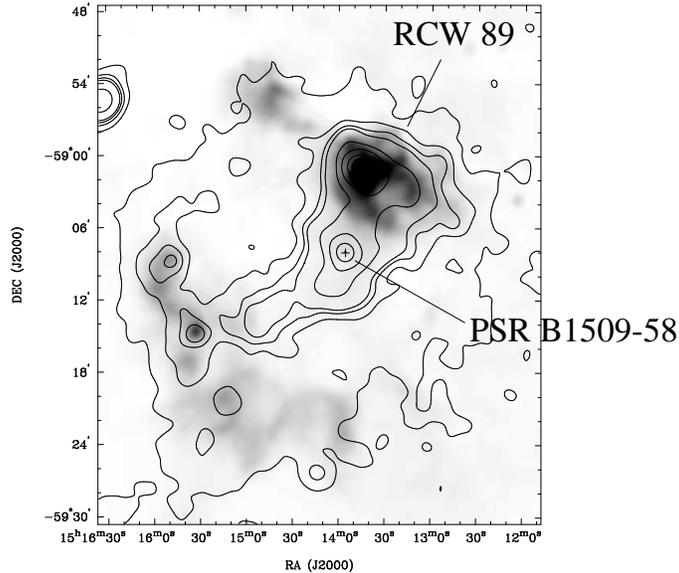,height=9cm,angle=270}}
\caption[h]{Radio and X-ray observations of \g320. The grey-scale
is a 36~cm MOST image (Whiteoak \& Green 1996), while the 
contours correspond to {\em ROSAT}\ PSPC data (Trussoni et al 1996).
The position of PSR~\psr\ is marked by a ``+'' symbol.}
\label{fig_most_rosat}
\end{figure}

Existing radio and X-ray images of \g320\ are shown in 
Figure~\ref{fig_most_rosat}. The radio image shows two distinct
components: a bright centrally-condensed source to the north
(coincident with the optical nebula \rcw), and a fainter filamentary
arc to to the south. In X-rays three bright sources are apparent:
the pulsar itself, a surrounding non-thermal pulsar-powered
nebula (a ``plerion''), and thermal emission from the \rcw\ region.

\subsection{Are \g320\ and PSR~\psr\ associated?}
\label{sec_assoc}

The distance of 4.2~kpc determined for \rcw\ through \HI\
absorption (Caswell et al 1975) is roughly consistent with
the pulsar's dispersion-measure distance of $\sim$6~kpc.
However the ages of the pulsar and SNR
are hard to reconcile:
assuming standard supernova and ISM parameters results in
an age for the SNR in the range $(6-21)\times10^3$~yr (Seward et al
1983), much older than the pulsar's spin-down age of $\la1700$~yr
(Kaspi et al 1994). 

On the other hand, an association is favoured by the possibility that
the pulsar may be interacting with the \rcw\ region through a
collimated relativistic outflow (Seward et al 1983; Manchester \&
Durdin 1983; Tamura et al 1996; Brazier \& Becker 1997). There are two
main problems with this argument, however.  Firstly, the total thermal
energy in \rcw\ exceeds the {\em entire}\ energy lost by the pulsar
over its lifetime for any but very rapid initial periods. Secondly, the
mechanism by which a large thermal nebula might be created at the
termination of such an outflow is unclear.

\subsection{What is \g320?}

\g320\ is not a typical, roughly circular, shell SNR.  Some of the possible
explanations for its unusual morphology are that it is actually
multiple remnants, that its appearance is the result of expansion into
a complicated environment, or that an outflow from the pulsar (see
\S\ref{sec_assoc}) has distorted its shape.

\section{New Observations and Results}

We have carried out extensive observations of the region with the
Australia Telescope Compact Array (ATCA). The main results are
as follows:

\begin{itemize}
\item \HI\ absorption is consistent
with all radio components of \g320\ being at a distance
\mbox{$5.2\pm1.4$}~kpc (PSR~\psr\ is too weak to obtain
\HI\ absorption against).
\item In the region surrounding PSR~\psr, radio emission from \g320\ 
is highly linearly polarised ($\sim$60\%) and
has the same rotation measure as the pulsar. As shown in 
Figure~\ref{fig_atca_pspc},
within
this plateau of emission is a channel of reduced radio
emission which closely follows
a collimated X-ray feature (part of the plerion powered
by PSR~\psr).
\begin{figure}
\centerline{\epsfig{file=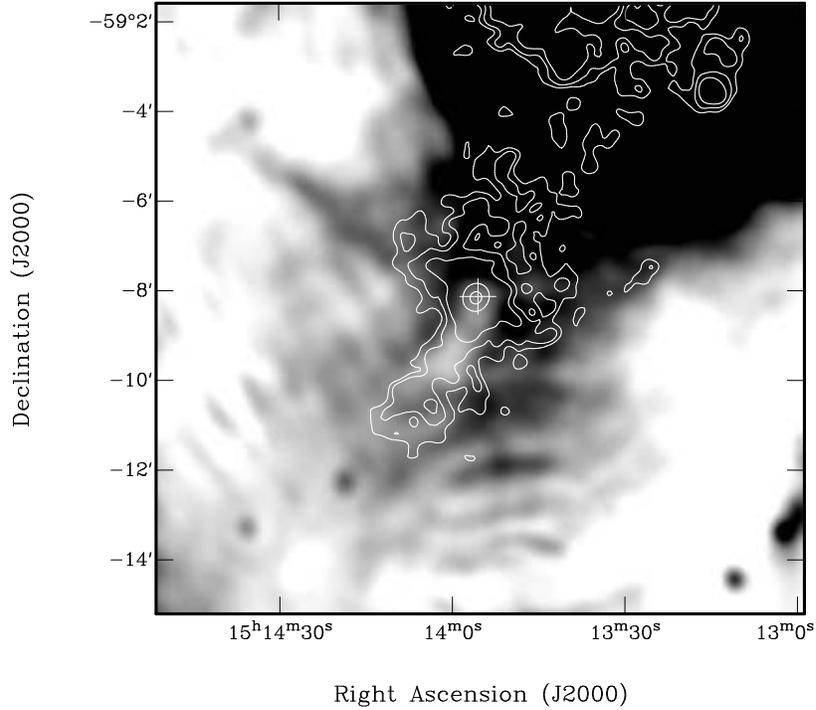,height=11cm}}
\caption[h]{Radio / X-ray comparison of the region near PSR~\psr. The
grey-scale corresponds to 20~cm ATCA data, while the contours 
represent {\em ROSAT}\ PSPC data (Greiveldinger et al 1995). 
The pulsar's position is marked
by a ``+''; corrugations seen in the radio emission 
are low-level artifacts resulting from deconvolution.}
\label{fig_atca_pspc}
\end{figure}

\item At high resolution, the radio peak of \rcw\ forms a ring of
linearly polarised knots of spectral index $\alpha \approx -0.5$ ($S_\nu
\propto \nu^{\alpha}$). The radio and X-ray morphologies
of the region are compared in
Figure~\ref{fig_hri}, where it can be seen that there is a marked
correspondence between the two wavebands.
\begin{figure}
\centerline{\epsfig{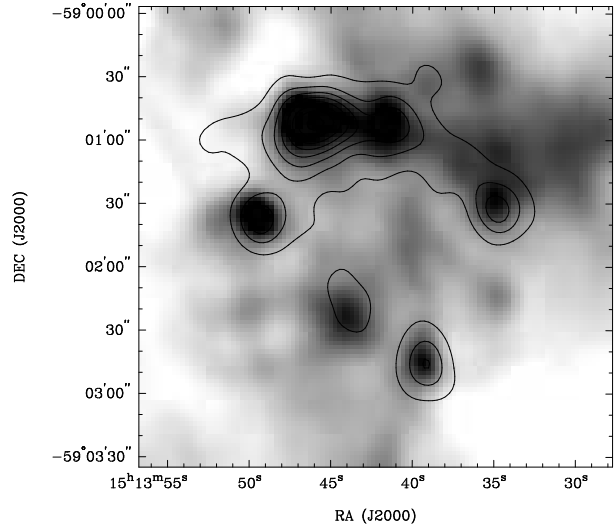}}
\caption[h]{Radio / X-ray comparison of the \rcw\ region. The
grey-scale corresponds to 6~cm ATCA data, while the contours 
represent smoothed {\em ROSAT}\ HRI data (Brazier \& Becker 1997).}
\label{fig_hri}
\end{figure}

\end{itemize}

\section{Discussion}

The radio emission seen near the pulsar in Figure~\ref{fig_atca_pspc}
has a distinctly high fractional polarisation, 
has the same rotation measure as the pulsar,
and shows an anti-correspondence with X-ray emission
associated with the plerion. Thus it seems certain that this radio
emission is in some way connected with the pulsar. We would argue that
this region is not radio emission from the plerion itself, since the
radio emission is not centred on the pulsar, nor are the X-ray bright
regions bright in the radio. Indeed any radio plerion is expected to be
too faint to be detectable (e.g.\ Seward et al 1984).
Figure~\ref{fig_atca_pspc} demonstrates that radio emission is enhanced
along the sides of the collimated X-ray feature but not within it. 
This anti-correspondence can be explained if the X-ray
feature is a focussed relativistic jet or outflow, as has been argued by
Brazier \& Becker (1997).  Radio emission is then produced by shocks
generated in a cylindrical sheath around the jet, as seen for SS 433
(Hjellming \& Johnston 1986, 1988) and possibly for the jet in Vela~X (Frail
et al 1997).

The polarisation and spectral index of the knots in
Figure~\ref{fig_hri} indicate that at least in the radio, their
emission mechanism is synchrotron. The X-ray / radio correspondence
suggests that the X-ray emission from these knots might also be
synchrotron, and indeed a single power law can be extrapolated from
their radio flux densities through to those measured by {\em
ROSAT}\ HRI. Although in X-rays the \rcw\ region as a whole is thermal,
a power-law spectrum from the knots could easily be hidden within. 

The problems raised in \S\ref{sec_assoc} involving an outflow from the
pulsar can be resolved if these compact synchrotron knots are
interpreted as the point of interaction between the pulsar outflow and
the SNR. The energy in these knots can be easily accommodated by the
pulsar spin-down, and it is easier to see how these features, rather
than the whole extended \rcw\ nebula, might be the termination of the
energetic outflow. The rest of the \rcw\ region may be part of the
SNR blast wave.

Correspondences between the discrete radio components of \g320\
and the overall X-ray remnant,
together with the results of \HI\ absorption, argue that
\g320\ is a single SNR. Its agreement in rotation
measure with PSR~\psr, and the evidence seen in Figures~\ref{fig_atca_pspc}
and \ref{fig_hri} for outflow and interaction respectively, make
a strong case that SNR~\g320\ and PSR~\psr\ are associated.

If the SNR and pulsar are indeed associated, then the age discrepancy
between them must somehow be resolved. We find that a consistent
picture of a system of age 1700~yr can be constructed if a supernova of
high energy or low mass ($E_{51}/M_{\rm ej} \sim 2$) occurred near the
edge of an elongated cavity. As well as the SNR's large apparent age,
this model can also explain the two well-separated components of the
remnant's radio morphology, the significant offset of the pulsar from
the SNR's centre and the faintness of any plerion at radio
wavelengths (see Gaensler et al 1998 for a detailed discussion of this
model).

\section{Conclusions}

Observations with the ATCA have resulted in the following main
conclusions:
\begin{enumerate}

\item The radio components of \g320\ are all part of a single
SNR at a distance $5.2\pm1.4$~kpc and with an age $\la$1700~yr.

\item PSR~\psr\ is physically associated with \g320. The pulsar emits
twin jets or collimated outflows of relativistic particles. One of
these jets is seen as a collimated X-ray feature surrounded by a
polarised radio sheath, while the other interacts with the SNR in the
form of radio/X-ray knots within \rcw.

\item SNR~\g320\ was formed in a supernova of high kinetic energy or low
ejected mass ($E_{51}/M_{\rm ej} \sim 2$) which occurred near the edge
of a low-density cavity.  
\end{enumerate}

A variety of further observations can support or refute these claims.
For example, the cavity we propose may be visible in \HI\ emission,
while forthcoming {\em AXAF}\ observations will allow a detailed study
of the physical conditions within the knots.  While we have
side-stepped the details of how pulsar jets might be
generated and how they interact with the SNR, it seems that PSR~\psr\
can now join the Crab, Vela and possibly PSR~B1951+32 in showing
evidence for collimated outflows; evidence is mounting that spherically
symmetric pulsar winds are a gross over-simplification.

\acknowledgements

BMG acknowledges the Local Organising Committee, the Science
Foundation for Physics within the University of Sydney, the R.\ and
M.\ Bentwich Scholarship and the James Kentley Memorial Scholarship for
financial assistance in attending this workshop. The Australia
Telescope is funded by the Commonwealth of Australia for operation as a
National Facility managed by CSIRO.


\end{document}